\documentstyle[aps,twocolumn,epsf,prl]{revtex}

\begin{document}

\twocolumn[\hsize\textwidth\columnwidth\hsize\csname 
@twocolumnfalse\endcsname
 
\title{On Ferromagnetism in the Large-$U$ Hubbard Model} 
 
\author{Thomas Obermeier, Thomas Pruschke, Joachim Keller}

\address{Institut f\"ur Theoretische Physik, Universit\"at Regensburg,
  Universit\"atsstr. 31, 93053 Regensburg, Germany}

\date{\today}
 
\maketitle
 
\begin{abstract}
We study the Hubbard model on a hypercubic lattice with
regard to the possibility of itinerant ferromagnetism. The Dynamical Mean Field
theory is used to map the lattice model on an effective local problem, which
is treated with help of the Non Crossing Approximation.
By investigating spin dependent one-particle Green's functions 
and the magnetic susceptibility, 
a region with nonvanishing ferromagnetic
polarization is found in the limit $U\to\infty$. The $\delta$-$T$-phase
diagram as well as thermodynamic quantities are discussed.
The dependence of the Curie temperature on the Coulomb interaction and the
competition between ferromagnetism and antiferromagnetism are studied in
the large $U$ limit of the Hubbard model.
\\

\\

\end{abstract}

]

The microscopic description of ferromagnetism in narrow-band metals
like Fe, Ni, Co and others is one of the most interesting problems
in solid state physics. Since the electrons in these systems are
mobile one cannot use localized-spin models with effective interactions
like e.g. the Heisenberg model, but has to take into account this
itineracy together with the electron-electron interaction on a more
fundamental level. The first model set up to describe such a system
is the Hubbard model \cite{Hub63}
\begin{equation}
        H=-t \sum_{<\rm{ij}>\sigma}{\rm c}^{\dagger}_{\rm i \sigma}
                {\rm c}_{\rm j \sigma} + U \sum_{\rm i} n_{\rm i \uparrow}
                n_{\rm i \downarrow}.
\end{equation}
However, it was realized relatively early that the Hubbard model rather
seems to be a generic model for antiferromagnetism \cite{And63} and 
a correlation driven metal-insulator transition instead. Just these
properties made it an early and rather successful candidate for
the description of the high-$T_c$ compounds \cite{And87}.

Nevertheless the question about ferromagnetism in the Hubbard model was 
never abandoned, since one of the few rigorous theorems about this model 
definitely proves its existence. In 1965 Nagaoka showed that for 
$U=\infty$ and one hole doped into the 
half-filled band the state with a fully polarized background is the
ground state for several lattice structures
due to a gain of kinetic energy for the hole \cite{Nag66}.
This theorem initiated a large amount of work on questions like the
stability of the Nagaoka state with respect to doping $\delta$, 
finite $U$, etc.
\cite{Lie93}. Moreover, even after 30 years of research the situation
appears to be rather controversial, especially for bipartite lattices:
One obtains critical dopings in the range $\delta_c$ from $0$ to $0.3$, 
depending on the method used \cite{Mue93,Put92}. Only for the infinite 
dimensional hypercubic lattice the situation seems to be clear:
The work by Fazekas et al. suggests that the Nagaoka state is unstable 
for any finite doping \cite{Faz90,Uhr96}, unless explicitely favoured by 
long-range Coulomb interactions \cite{Str95} or band structure effects
\cite{Ulm96}.

Most of the above studies are based on a variational ansatz and are thus
restricted to the Nagaoka - i.e. fully polarized - state and its stability
as the ground state. There is still the possibility of partially
polarized ferromagnetism and in any case the necessity to calculate
$T_c$ as function of $\delta$ and $U$ etc. Generally speaking the question
to what extent
ferromagnetism is a generic feature of the Hubbard model or not
is still unanswered. 

In this letter we discuss the magnetic phase diagram of the Hubbard model
on a hypercubic lattice for large Coulomb repulsion $U$. In the latter limit
the ground-state and low-energy properties of the model are well
captured by a $t$-$J$ model with an effective antiferromagnetic
exchange $J=2t^2/U$ \cite{tJmodel}. To solve the $t$-$J$ model or, more precisely,
the underlying Hubbard model at $U=\infty$ we use the dynamical mean field
theory (DMFT) \cite{Met89}. This theory leads to purely local
dynamical renormalizations of one-particle properties, which can be 
obtained from an
effective impurity problem coupled to a self-consistent medium \cite{Pru95,Geo96}.

In addition to the 
one-particle properties the DMFT also allows to calculate 
two-particle correlation functions and thermodynamic quantities 
consistently \cite{Jar93,Pru96}. Especially in the limit of large $U$
one obtains \cite{Pru96}
\begin{equation}\label{chi_F}
\chi_{F}^U(T) = \chi_{F}^\infty(T)/\left(1+ 2d\cdot\frac{2t^2}{U}
\chi_{F}^\infty(T)\right)
\end{equation}
and 
\begin{equation}\label{chi_AF}
\chi_{AF}^U(T)=\chi_{AF}^\infty(T)/\left(1-2d\cdot\frac{2t^2}{U}\chi_{AF}^\infty(T)\right)
\end{equation}
for the homogenous ($\chi_F^U(T)$) and staggered ($\chi_{AF}^U(T)$) susceptiblities
of the Hubbard model,
respectively. The quantity $\chi_\alpha^\infty(T)$ denotes the 
susceptibility for $U=\infty$ and $d$ is the spatial dimensions of the system.
These expressions allow to discuss the influence
of finite $U$ once the $\chi_\alpha^\infty(T)$ are known.

In the follwing we use $4dt^2=1$ as energy unit. With this choice
the bare density of states for the hypercubic lattice is of Gaussian form for
large $d$: $\rho_0(\varepsilon)=1/\sqrt{\pi}\exp(-\varepsilon^2)$.
The effective impurity problem of the DMFT
is solved within the NCA \cite{Kei70,Bic85} and for $U=\infty$ we furthermore
do the calculations with spin-dependent quantities to explicitely look at the
properties in the symmetry-broken phase. An extension of these
calculations to finite $U$ is extremely tedious and studies along this line
are in progress \cite{Obe97}.

\begin{figure}[!]
\begin{center}
\unitlength1cm
\begin{picture}(12,5.5)
\put(0,0.25){\epsfxsize=8.5cm\epsfbox{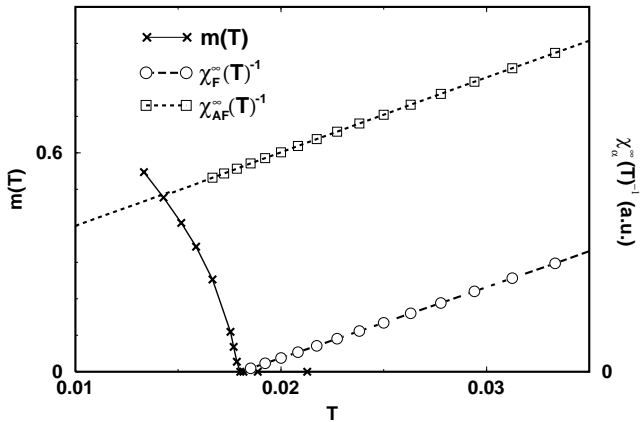}}
\end{picture}
\caption[]{Magnetization $m(T)$ (crosses), inverse homogenous susceptibility 
$\chi_F^\infty(T)^{-1}$ (circles) and staggered susceptibility 
$\chi_{AF}^\infty(T)^{-1}$ (squares) for $\delta=0.03$ as function of temperature.
}
\label{fig_moft}
\end{center}
\end{figure}
\begin{figure}[!]
\begin{center}
\unitlength1cm
\begin{picture}(12,5.5)
\put(0,0.05){\epsfxsize=8.5cm\epsfbox{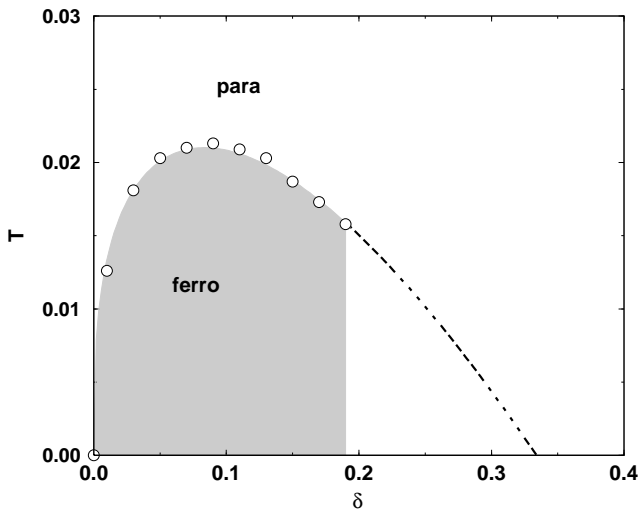}}
\end{picture}
\caption[]{Magnetic phase diagram for the Hubbard model on a hypercubic
lattice. The dashed line represents a fit to the last data points
and predicts a critical doping $\delta_c\approx0.33$ beyond which $T_C=0$.}
\label{fig_phas}
\end{center}
\end{figure}
We begin the discussion of our results with the case $U=\infty$. Figure~\ref{fig_moft}
shows the inverse susceptibility (homogenous and staggered) as function of 
temperature $T$ for a doping $\delta=1-\langle n\rangle=0.03$. While $\chi_{AF}^\infty(T)^{-1}$
remains finite for all $T$, $\chi_{F}^\infty(T)^{-1}$ vanishes linearly for a
$T_C>0$ and below $T_C$ we observe a finite magnetization
$m=n_{\uparrow}-n_{\downarrow}$ with $m(T)=c\sqrt{|T-T_c|}$.
Note that the critical points found from $m(T)$ and $\chi_F^\infty(T)$ conincide
indicating a second order transition (see Fig.~\ref{fig_moft}). Unfortunately
our data are not sufficient to extrapolate for $m(T=0)$, which according
to the results by Fazekas et al.\ \cite{Faz90} we would expect to have a value
$m(T=0)<n$. Clearly, this point needs further investigation.

Repeating the above calculation for different dopings we obtain the 
$\delta$-$T$-phase diagram in Fig.~\ref{fig_phas}, which shows a  
fairly extended region of ferromagnetism with a 
maximum in $T_c$ at $\delta$ between  $0.07$ and $0.08$.
Although the NCA in principle does not allow to do calculations down to $T_c$
beyond $\delta=0.1$, the observed Curie-Weiss form of $\chi_F^\infty(T)$
enables us to obtain data points in this region of the phase diagram
from $\chi_F^\infty(T)$ at high temperatures.
Obvioulsy this procedure becomes less accurate for increasing doping so that
the behaviour of the phaseline $T_c(\delta)$ currently remains unknown
for $\delta>0.2$.
The extrapolation of the available data nevertheless indicates that a critical
doping $\delta_c$ between $0.3$ and $0.4$ exists beyond which $T_C=0$.

\begin{figure}[!]
\begin{center}
\unitlength1cm
\begin{picture}(12,8.5)
\put(0,0.25){\epsfxsize=8.5cm\epsfbox{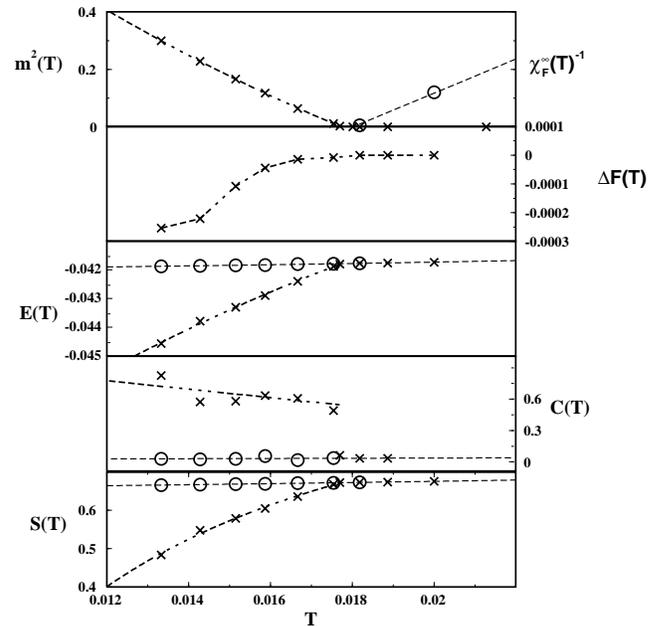}}
\end{picture}
\caption[]{Squared magnetization $m(T)^2$, inverse homogenous susceptibility
$\chi_F^\infty(T)^{-1}$, difference in free energies
$\Delta F(T) = F_{\rm FM}(T)-F_{\rm PM}(T)$, internal energy $E(T)$, specific
heat $C(T)$ and entropy $S(T)$ for $\delta=0.03$ in the paramagnetic (circles)
and ferromagnetic phase (crosses) close to $T_C$. From $\Delta F(T<T_C)<0$ it
is clear that below $T_c$ the ferromagnetic solution is stable.}
\label{fig_tdyn}
\end{center}
\end{figure}

The stability of the ordered phase depends on the interplay of internal
energy $E(T)$ and entropy $S(T)$ entering the free energy
$F=E-TS$.
In Fig.~\ref{fig_tdyn} we thus show the difference in free
energies $\Delta F(T)=F_{\rm FM}(T)-F_{\rm PM}(T)$ together with the internal energy,
specific heat and  entropy for the ferromagnetic and paramagnetic state.
To relate the data to the preceeding discussion the magnetization and inverse
susceptibility from Fig.~\ref{fig_moft} are shown again in the upper part of 
Fig.~\ref{fig_tdyn}. Below $T_C$ the difference $\Delta F(T)$ becomes negative,
i.e.\ the ferromagnetic state is indeed thermodynamically stable.

The internal energy $E(T,n)$ is given by the expectation value of $H$, which
for $U=\infty$ is equivalent to the kinetic energy. As is evident from
Fig.~\ref{fig_tdyn}
$E(T)$ for the ferromagnetic solution is lower than the corresponding values
in the paramagnet below $T_c$. Therefore
the transition to the ferromagnetic phase is obviously connected
with a gain in kinetic energy. This leads to the conjecture
that the  physics underlying the stability of the ferromagnetic state
should be roughly the same as in the particular case studied within the
Nagaoka theorem. Since at $T_c$ the slope of the internal energy $E(T)$ changes
for the ferromagnetic solution the specific heat $C(T)=\partial E/\partial T$
shows a jump characteristic for a second order phase transition. Note that only a very small
temperature region around $T_C$ is shown in Fig.~\ref{fig_tdyn} and that $C(T)$
decreases again for lower temperatures.
Finally, the entropy $S(T)$ is obtained from 
$
F=E-TS
$.
Just above $T_C$ its value is very close to $\ln{2}$, the value expected for a
spin $1/2$ system,while below $T_C$ the increasing spin order
leads to a strong decrease of $S(T)$.

\begin{figure}[!]
\begin{center}
\unitlength1cm
\begin{picture}(12,6)
\put(-8.5,0.25){\epsfxsize=6.5cm\epsfbox{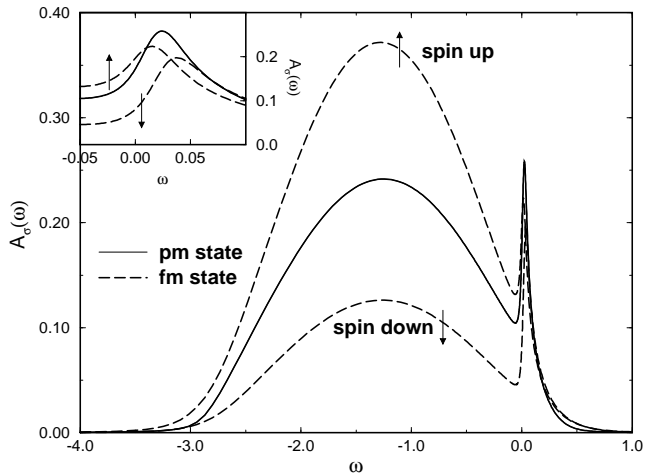}}
\end{picture}
\caption[]{
Spectral functions for $\delta=0.03$ and $\beta=70$ for both spins.
The temperature is well below $T_c$, so that the spin dependent
solution (dashed line)  shows a difference in the spectral weight
for the two spin directions.
The inset shows details near the Fermi energy.
}
\label{fig_GF}
\end{center}
\end{figure}
The ordered state of course also shows up in the dynamical
properties such as the one-particle Green's function. In Fig.~\ref{fig_GF}
we show the density of states (DOS) for a doping $\delta=0.03$ at a temperature
$T=1/70<T_C$ for both the paramagnetic (full line) and ferromagnetic solution
(dashed lines).
The basic features in the ferromagnetic phase
are similar to those of the paramagnet. 
One finds the lower Hubbard band represented by a broad peak and 
a quasi-particle resonance near the Fermi energy \cite{Pru95,Geo96}. 
Due to $U=\infty$ the upper Hubbard band does not appear.

Depending on temperature and doping 
spectral weight is transferred between the states with spin $\sigma$
and that of $-\sigma$, most prominent in the charge-fluctuation peaks, which 
results in differences in the
occupation numbers $n_{\sigma}$ and
in a finite magnetization $m=n_{\uparrow}-n_{\downarrow}$.
Note that this does not occur due to any explicit magnetic exchange
but rather to the fact that the energy loss by increasing the
population of the $\sigma$-band is outweighed by the gain in kinetic
energy from the holes in the $-\sigma$-states \cite{Nag66}. In addition
the peak positions for the minority/majority spins are shifted to a
somewhat higher/lower energy.
In terms of a band picture this means a slight splitting of 
the lower Hubbard band.

The energy splitting is observed for the quasi-particle resonance
near the Fermi energy as well (see inset of Fig.~\ref{fig_GF}).
In contrast to the lower Hubbard band {\it both peaks}
show a loss of spectral weight
compared to the para\-magnetic state.
This reflects the suppression of the Kondo like effect underlying the
quasi-particle resonance by ferromagnetism, analogous to the effect of
an external magnetic field in conventional Kondo physics.
With increasing magnetization the resonances will continuoulsy decrease in
height and eventually vanish for very low temperatures.

\begin{figure}[!]
\begin{center}
\unitlength1cm
\begin{picture}(12,7)
\put(0,0){\epsfxsize=8.5cm\epsfbox{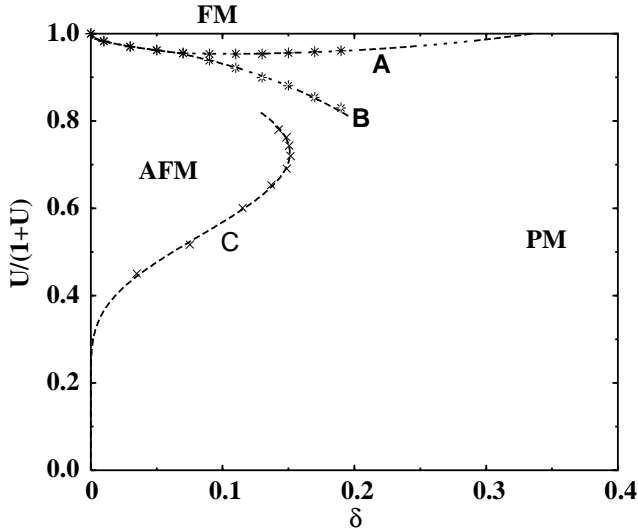}}
\end{picture}
\caption[]{$\delta$--$U$ phase diagram of the Hubbard model from the large $U$
limit (curves {\sf A} and {\sf B}). We find a region with ferromagnetism above 
curve {\sf A}. Below curve {\sf B} antiferromagnetism is predicted.
The phase line at small $U$ (curve {\sf C}) is taken from \cite{Jar93} and
extrapolates to our phase line for $U\to\infty$.
}
\label{fig_pdf}
\end{center}
\end{figure}

Let us now turn to the interesting question of the dependence of these results
on $U$. Generally speaking we expect that with decreasing $U$ the Curie
temperature $T_C$ should be supressed and, due to the antiferromagnetism favoured
by a finite $U$, a competition between ferromagnetic and antiferromagnetic order
should occur. This anticipated behaviour can readily be read off the signs in
equations
(\ref{chi_F}) and (\ref{chi_AF}), i.e.\ $U<\infty$ tends to supress $\chi_F^U(T)$
and enhance $\chi_{AF}^U(T)$.
In addition, the Curie-Weiss form of $\chi_F^U(T)$ and $\chi_{AF}^U(T)$ (see
Fig.~\ref{fig_moft}) allows to rewrite eqs.\ (\ref{chi_F}) and (\ref{chi_AF})
in such a way as to identify a Curie temperature
$T_C(U,\delta)=T_C^\infty(\delta)-C(\delta)/U$ and a N\'eel temperature
$T_N(U,\delta)=-\Theta(\delta)+\tilde{C}(\delta)/U$,
where $T_C^\infty(\delta)$ and $C(\delta)$ are the Curie temperature and
Curie constant for $U=\infty$, while $\Theta(\delta)$ and $\tilde{C}(\delta)$
denote the intercept and inverse slope of $\chi_{AF}^\infty(T)^{-1}$.
A detailed discussions of $C(\delta)$, $\tilde{C}(\delta)$ and $\Theta(\delta)$
will be given elsewhere \cite{Obe97}. Here we want to focus on the resulting
$\delta$-$U$ phase diagram in Fig.~\ref{fig_pdf}, curves {\sf A} and {\sf B},
which were obtained by plotting $\max(T_C(U,\delta),T_N(U,\delta),0)$.

One sees that for large $U$ an extended region of ferromagnetism exists
above curve {\sf A}, which is completely supressed for $U<U_c\approx20$.
For decreasing $U$ the ferromagnetic order is eventually
replaced by antiferromagnetism in the region below curve {\sf B}. Note that up to
$\delta\approx0.07$ we find a direct transition from the ferromagnet to the
antiferromagnet, which we would expect to be of first order ending in a second
order critical point. A more detailed investigation of the region would thus
be of great interest. However, since the transition temperatures are already
very small there we do not see any chance to achieve this with the methods
currently available. Beyond $\delta\agt0.07$ a paramagnetic region separates the
two phases.
In addition to our new findings we also include results on the phase
line between antiferromagnet and paramagnet for the full Hubbard model (1) at
small $U$ (curve {\sf C}) \cite{Jar93}. The behaviour for the largest $U$ values in this case
extrapolates nicely to our phase line {\sf B} for $U\to\infty$.
For decreasing
$U$ and increasing $\delta$, however, the approximation of the Coulomb term by
an effective exchange becomes worse, i.e.\ the magnetic order is
much stronger supressed by doping for a given $U$.

To conclude, we have shown that the Hubbard model on a hypercubic lattice
provides a scenario of ferromagnetism for finite doping 
and at finite temperatures. We were for the first time able to obtain
sensible results for $T_C$ as function of $\delta$ and $U$ for the strong
coupling case. As in Nagaoka's case the phase transition originates from a
gain of kinetic energy, as we could see from thermodynamic quantities. 
The $\delta$-$T$-$U$ phase
diagram shows a fairly extended region of ferromagnetism for large $U$ that
is completely suppressed for $U<U_c\approx20$ and $\delta>\delta_c\approx0.3$.
For small doping and large $U$ we observed in addition a direct transition from
the ferromagnet into an antiferromagnet as function of $U$.

Unfortunately our method to solve the DMFT does not allow to study
temperatures $T\ll T_c$. Thus several important questions have to remain unanswered:
What is the ground-state magnetization $m(T\to0)$ (cf.\ \cite{Faz90,Uhr96}) and
of what nature is the ferromagnetic $\leftrightarrow$ antiferromagnetic
transition, for example. In future work one also must investigate the order of
the transition paramagnet $\leftrightarrow$ ferromagnet,
which is under current discussion (cf. ref. \cite{Her96}, where
a first order transition is stated within a different method), more closely.

We thank Ch.~Helm and W.~Heindl for helpful discussions.
This work was supported by the Deutsche Forschungsgemeinschaft grant
number Pr 298/3-1.

\end{document}